\documentclass[aps,prl,twocolumn,showpacs]{revtex4}

\usepackage[dvips]{graphicx}
\usepackage{epsfig}
\usepackage{amsmath}
\usepackage{amsfonts}
\usepackage{amssymb}
\usepackage{wasysym}

\expandafter\ifx\csname package@font\endcsname\relax\else
 \expandafter\expandafter
 \expandafter\usepackage
 \expandafter\expandafter
  \expandafter{\csname package@font\endcsname}%
\fi

\begin{document}
\def\be{\begin{equation}}
\def\ee{\end{equation}}

\title{Dynamical scaling in branching models for seismicity}

\author{Eugenio Lippiello,$^1$ Cataldo Godano$^2$ and Lucilla de Arcangelis$^3$}
\affiliation{$^1$ Department of Physical Sciences,
University of Naples "Federico II", 80125 Napoli, Italy \\
$^2$ Department of Environmental Sciences and CNISM, Second
University of Naples, 81100 Caserta, Italy \\
$^3$ Department of
Information Engineering and CNISM, Second University of
Naples, 81031 Aversa (CE), Italy}

\begin{abstract}

We propose a branching process based on a dynamical scaling
hypothesis relating time and mass.
In the context of earthquake occurrence, we show that
 experimental power laws in size and time distribution
naturally originate solely from this scaling hypothesis.
We present a numerical protocol able to generate
a synthetic catalog with an arbitrary large number of events.
The numerical data reproduce the hierarchical organization in time
and magnitude  of experimental inter-event time distribution.

\end{abstract}

\pacs{02.50.Ey,64.60.Ht,89.75.Da,91.30.Dk}

\maketitle

Stochastic branching models have been used since a long time in
the description of a large variety of social and physical
phenomena ranging from biological evolution or genealogy 
\cite{branching}, to nuclear or chemical reactions
\cite{branching1}. In the last years branching processes have been
widely studied in seismicity and actually they provide one of the most
efficient tools for earthquake forecasting \cite{gerg}. Within this
approach, one treats seismicity as a marked point process in time
$\{M_i(t_i)\}$ where $t_i$ is the occurrence time of an event with
mass (magnitude) $M_i$ and one assumes that each event can trigger
future ones according to a two point conditional rate
$\rho(M(t) \vert M_i(t_i))$. Given a history of past events
$\{M_i(t_i)\}$, then, the rate of events of magnitude $M$ at time
$t$ is given by \be \rho(M(t) \vert \{M_i(t_i)\})= \sum _{i:t_i<t}
\rho(M(t) \vert M_i(t_i))+\mu P(M)
    \label{VJ}
    \ee
where $\mu$ is a constant rate of independent sources 
and $P(M)$  their magnitude distribution. In the epidemic type
aftershock sequences (ETAS) model \cite{Ogata}, widely used in
seismology, for any couple of events $i$ and $j$, magnitude 
of event $i$ is independent of previous events
\be
    \rho(M_i(t_i) \vert M_j(t_j))=P(M_i)g(t_i-t_j;M_j)
    \label{FAC}
    \ee
and for the magnitude distribution $P(M_i)$ and the propagator
$g(t_i-t_j;M_j)$ one uses three well known experimental observations:

i) The magnitude distribution follows an exponential law
$P(M)\sim 10^{-bM}$ usually referred as the Gutenberg-Richter (GR)
law \cite{gutri}, where $b\simeq 1$;

ii) The Omori law \cite{Omori} states that the number of
aftershocks $n(t)$ decays in time as $n(t)\sim (t+c)^{-p}$
with $p\simeq 1$;

iii) The aftershock number is exponentially related to the mainshock
magnitude. This last law  combined with the Omori law gives
$g(t_i-t_j,M_j)\propto 10^{\alpha M_j}(t_i-t_j+c)^{-p}$, where $\alpha\sim b$
\cite{helm}.

The main statistical properties of the ETAS model have been reviewed
in a series of papers (see for instance ref. \cite{sornette}). Ultraviolet
and infrared cut-offs have to be necessarily introduced
to make the theory convergent.  Whereas a large magnitude
cut-off can be expected on physical grounds, more questionable is the
existence of a minimum magnitude $M_{inf}$ \cite{werner}.
This problem has been
removed by a self-similar version of the ETAS model recently
introduced by Vere-Jones \cite{vere}. In this model the
decoupling (\ref{FAC}) between magnitude and time is still assumed but a
multiplicative factor is considered
$\rho(M_i(t_i) \vert M_j(t_j))=P(M_i)g(t_i-t_j;M_j)S(M_i-M_j)$
where $S(M_i-M_j)=10^{-d \vert M_i -M_j \vert}$
introduces magnitude correlations: for $d>0$ large
daughter earthquakes tend to occur after large mother earthquakes.
Saichev and Sornette have suggested a physical
explanation for this magnitude
correlation based on faults branching \cite{saichev}.

In this paper we show
that the above mentioned magnitude correlations
do not need to be introduced via an ad hoc term, as $S(M_i-M_j)$, but
naturally originate from a more general scaling relation. More
precisely we assume that the
magnitude
difference $M_i-M_j$ fixes a characteristic time scale
\be
\tau_{ij}=k 10^{b(M_j-M_i)}
\label{tau}
\ee
so that the conditional rate is magnitude independent when time is rescaled by
$\tau_{ij}$ and $k$ is a constant measured in seconds
\be
    \rho(M_i(t_i) \vert M_j(t_j)) =F \left [
    \frac{t_i-t_j}{\tau_{ij}}\right ] .
    \label{scaling}
    \ee
With the only constraint that $F(x)$ can be normalized, we 
recover the statistical features of earthquake
occurrence: Omori law, GR law and scaling behaviour of the interevent
time distribution \cite{Corral,Bak}. Furthermore, using 
Eq.s (\ref{VJ},\ref{scaling}) in a 
numerical code we are
able to generate, in few hours of CPU time, a synthetic catalog with
the same number of events as 30 years California catalog. Experimental
and numerical catalogs are found to exhibit the same time and magnitude
organization.

In order to verify the existence of magnitude correlations we consider the
ANSS California catalog \cite{cali} containing $N=9586$ $M>3$ earthquakes.
We divide the catalog in $N_L=N/L$ subsets each containing $L=125$
events and define the quantity
$\delta M_j=(1/L)\sum_{i=j*L+1}^{j*L+L}M_i-(1/N)\sum_{i=1}^{N}M_i$,
representing the deviation of the average magnitude in
the $j$-th subset with respect to the average over the entire catalog.
We then calculate the quantity $C(n)$ defined as
\be
C(n)={1\over l} \sum_{j=n+1}^{n+l} \frac{1}{N_L-n_{max}}
\sum_{i=1}^{N_L-n_{max}}\delta M_i\delta M_{i+j}
\ee
where $n_{max}=32$ is the maximum "distance" between subsets considered.
The advantage of correlating average quantities $\delta M_j$ is
to reduce fluctuations and the
sum over $j$ further smoothens statistical noise. 
In absence of magnitude correlations,
$\delta M_i$ and $\delta M_{i+j}$ are statistically independent
quantities and $C(n)$ does not depend on $n$ and
fluctuates around $C(n)=0$. This situation can be realized by evaluating
$C(n)$ after reshuffling the magnitudes. Using $100000$ realizations
of the reshuffled catalog, we find that the distribution of
$C(n)$ exhibits gaussian behaviour centered in zero with standard
deviation $\sigma=0.0001$.  Figure 1, conversely, shows
that $C(n)$ computed for the real catalog has a regular trend with
amplitude several times larger than $\sigma$,
clearly indicating the existence of magnitude
correlations.
Therefore the behaviour of $C(n)$ for the real catalog is
a signature of a well defined earthquake magnitude organization.

\begin{figure}
\includegraphics[width=8cm]{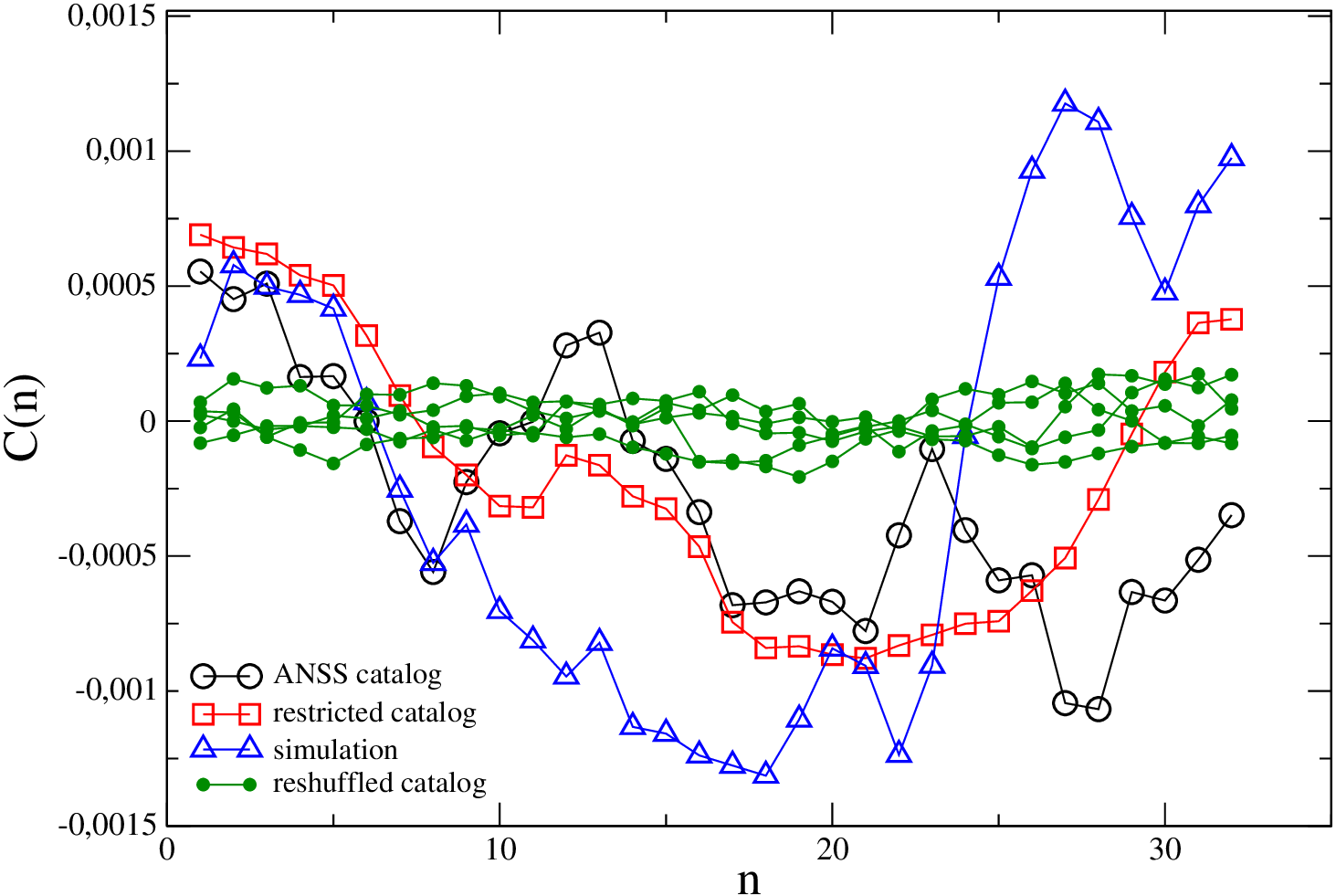}
\caption{(Color online) Function $C(n)$ versus $n$  for the ANSS catalog 
($\circ$), the restricted catalog ($\square$) and the simulated catalog
($\triangle$). $C(n)$ evaluated for five different realizations of 
reshuffled ANSS catalog (dots).} 
\label{fig1}
\end{figure}

In order to check that the results of Fig. 1 are not a spurious effect of
short term aftershock incompleteness \cite{Kag},
we use the method proposed by Helmstetter et al
\cite{helm} stating that, after a main shock of magnitude $M_M$ at time $t_M$,
the completeness level $M>3$ is recovered only after a time
$t_C=t_M+10^{(M_M-7.5)/0.75}$. We then construct a restricted catalog by
neglecting all events occurring within a time interval $[t_M,t_C]$
after each event with $M>1$, obtaining a catalog complete for $M>3$ and
   containing  $N=8502$ events.
The evaluation of $C(n)$  (Fig.1) now shows that correlations do not
disappear but
$C(n)$ exhibits again a regular trend with values significantly different than
zero.  The comparison of $C(n)$ in the original and restricted catalog
indicates that magnitude correlations are not therefore an artifact of
catalog incompleteness but should be attributed to a physical effect due to
earthquake interactions.
These interactions are
introduced in our approach  by means of the  scaling assumption
(\ref{scaling}). In this way we are able to reproduce, at least at a
qualitative level,   the experimental behaviour of $C(n)$  as can be
seen in Fig.1. 

Before discussing the details of our numerical procedure, we explore 
the consequences of our scaling assumption (\ref{scaling}). 
We first notice that the total number of daughter earthquakes, conditioned 
to the occurrence of a mother earthquake of
   magnitude $M_0$ at time $t_0$, is given by
$\int _{t_0}^\infty dt \rho(M(t)\vert
   M_0(t_0))=k 10^{-b(M-M_0)} \int_0^\infty dx F(x)$. Since $F(x)$ is
normalizable, we recover the GR behaviour independently of the specific
form of $F(x)$. On the basis of this observation, a
mother earthquake at time $t_0$ will be distributed according to GR law
and therefore
the occurrence rate of magnitude $M$ triggered events at time $t$
is then given by $\rho(M,t-t_0)=\int_{M_{inf}}^{M_{sup}}
\rho(M(t)\vert M_0(t_0)) P(M_0)dM_0$ which leads to
\be
    \rho(M,t-t_0) \propto \frac{10^{-b M}}{(t-t_0)} \int_
    {10^{-b(M_{sup}-M)(t-t_0)}}^{10^{b(M-M_{inf})(t-t_0)}}
    F(z)dz
    \label{Pm1}
    \ee
We first observe that ultraviolet and
infrared cut-offs are not necessary anymore.  Indeed, in Eq.(\ref{Pm1})
one can arbitrary set $M_{inf}\to -\infty $ and
$M_{sup}\to \infty$ obtaining $
\rho(M,t-t_0) \propto \frac{10^{-b M}}{(t-t_0)}$. 
Then one recovers, as a direct consequence of the only
assumption (\ref{scaling}), beside the GR also the Omori law that
conversely are assumed a priori in the ETAS model.
The above result, that hold for quite arbitrary $F(x)$,
suggests that these two fundamental laws, generally considered as
independent laws in seismicity, can be strictly related to a more
general scaling behaviour.
 Furthermore
Eq.(\ref{Pm1}) shows that a change in $M_{sup}$ or $M_{inf}$ only
corresponds to a time rescaling and therefore the only effect is on
 the parameter $k$ in (\ref{tau}).

We now construct with our approach a
synthetic catalog to be compared with  the experimental one.
In a numerical protocol one assumes at initial time $t_0=0$ a
single event of arbitrary magnitude chosen in a fixed range
$[M_{inf},M_{sup} ]$. Time is then increased by a unit step
$t=t_0+1$, a trial magnitude is randomly chosen in the interval
$[M_{inf},M_{sup} ]$ and Eqs.(\ref{VJ},\ref{scaling})
give the probability to
have an earthquake in the time window $(t_0,t_0+1)$. If this
probability is larger than a random number between $0$ and $1$, an
earthquake takes place, its magnitude and occurrence time are
stored and used for the evaluation of probability for
future events. Time is then increased and in this way one
constructs a synthetic catalog of $N_e$ events. The term $\mu$ in
Eq.(\ref{VJ}) represents an additional source of earthquakes
Poissonian distributed in time with a magnitude chosen from the GR
distribution with $b=0.8$. Other choices for $P(M)$ do not sensibly affect 
the statistical properties of the simulated catalog.

Following this protocol, we generate sequences of $15000$ events
using a power law form for $F(z)=A/(z^\lambda+\gamma)$
with $M_{inf}=1$ and $M_{sup}=8$.
We compute magnitude distribution $P(M)$ and intertime
distribution  $D(\Delta t,M_L)$
where $\Delta t$ is the time distance between successive
events with
magnitude greater than a given threshold $M_L$. Extended analysis
of experimental catalogs have shown \cite{Corral,Bak} 
that the intertime distribution
is a fundamental quantity to characterize the magnitude and time
organization of earthquakes. In fact,
indicating with
$P_C(M)$ the cumulative magnitude distribution inside the
considered region, one observes
\be
D(\Delta t,M_L)=P_C(M_L) f(P_C(M_L)\Delta t) \label{cor}
\end{equation}
where $f$ is a universal function, independent on $M_L$ and on
the geographical region indicating a well defined hierarchical
organization of earthquake occurrence \cite{Corral2}. 
The numerical  distributions are compared
with the experimental data from the ANSS Catalog.
For different values of $\lambda$, it is always possible to find a set of
parameters ${A,\gamma,b,\mu}$ such that numerical data reproduce, on average,
earthquake statistical features both in time and in
magnitude. The parameter $k$ is fixed a posteriori in order to obtain
the collapse between numerical and experimental time.
We have also performed simulations for different values of $M_{inf}$
and $M_{sup} $ obtaining similar results and confirming that
changes in the magnitude range only produce time rescaling.

In Fig.2 we plot the experimental and numerical $D(\Delta t,M_L)$
considering two different values of $\lambda$
$(\lambda=1.2$ and $5)$ and $M_L$ ($M_L=1.5$ and $2.5$).
Data for different values of the parameters follow a universal curve and the
same collapse is obtained for other values of $\lambda>1$.
The accordance between experimental and numerical curves 
(inset of Fig.2) indicates that
the hypothesis of dynamical scaling is able to reproduce
two fundamental properties of seismic occurrence, namely the GR law
and Eq. (\ref{cor}), independently of the details of $F(z)$.

The ETAS model corresponds to a particular choice for $F(z)$, i.e. 
$\gamma=0$ and $\lambda =p\simeq 1$. We want to stress the important
difference due to the presence of a non-zero $\gamma$. First,
the constant $\gamma$ removes the problematic
need of an infrared (ultraviolet) cut-off in the ETAS model, whereas
in our approach  $M_{inf}$  and  $M_{sup}$ are irrelevant variables. 
 Second, the constant $\gamma$ gives
rise to the observed magnitude correlation. Indeed, for a given
mainshock of magnitude $M_j$ at time $t_j$, at each time
$(t_i>t_j)$ it is possible to define a sufficiently large
magnitude difference
$\Delta M$ such that, if $M_j - M_i > \Delta M$, we have
that $z^\lambda$ is negligible with respect to $\gamma$ and therefore
$F[(t_i-t_j)/\tau_{ij}]\simeq A/\gamma$. In other words after a large
event, the probability of big quakes is raised.

\begin{figure}
\includegraphics[width=8cm]{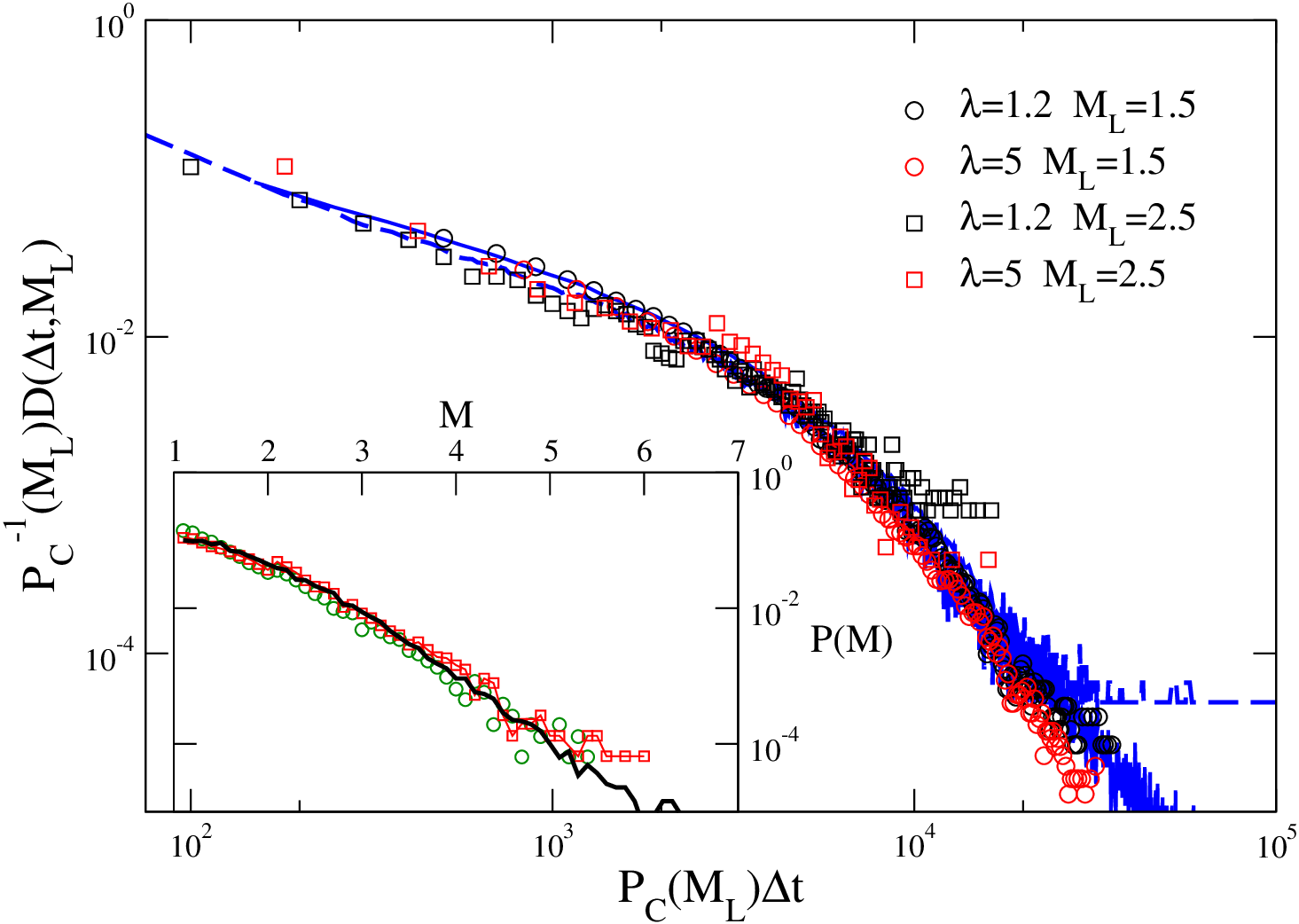}
\caption{(Color online)
The intertime distribution with  $F(z)=A/(z^\lambda+\gamma)$,
with two values of $\lambda=1.2,5$ and $M_L=1.5,2.5$.
Continuous and broken curve are the experimental $D(\Delta t,M_L)$
with $M_L=1.5$ and $M_L=2.5$.
For $\lambda=1.2$ ($\lambda=5$) we set $k=210sec$ ($k=420sec$), 
$A=1.4\ 10^{-4}sec^{-1}$ ( $A=1.9\ 10^{-4}sec^{-1}$), $\mu =4
 \ 10^{-7}$ ($\mu =1.5 10^{-6}$), $\gamma=1$ ($\gamma=0.1$)
and $b=1$.
In the inset the
magnitude distribution of the experimental (black line)
and numerical catalog with  $\lambda=1.2$ (red $\circ$) and
  $\lambda=5$ (green $\square$).
}
\end{figure}

\begin{figure}
\includegraphics[width=8cm]{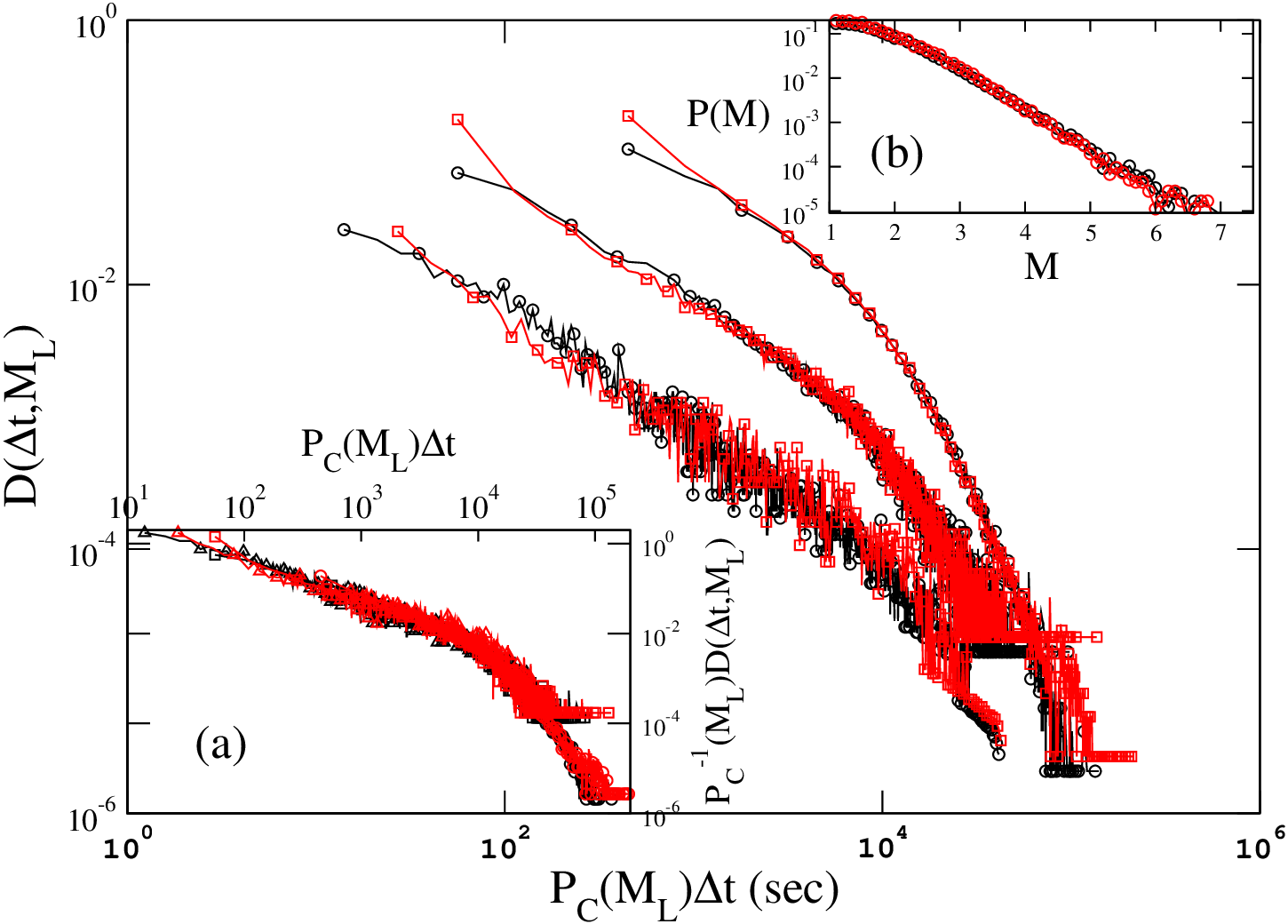}
\caption{(Color online) The intertime distribution as function
of $\Delta t P_c(M_L)$
obtained using Eq.(\ref{fz2}) (black circle {\Large $\circ$}) and
compared with the experimental distributions (red $\square$)
for three different
values of $M_L$ ($M_L=1.5,2.5,3.5$, from top to bottom).
We set $k=4.9\ 10^4 sec$, $A=6.1\ 10^{-5}sec^{-1}$, $\mu =2\ 10^{-5}$,
$\gamma=0.1$.
In inset (a) collapse of the three curves for different $M_L$ following 
the scaling of Eq.(\ref{cor}) and in inset (b) the experimental (red) and
numerical (black) magnitude distribution.
} \label{fig3}
\end{figure}

We have also performed more extensive simulations using for $F(z)$
an exponential behaviour
\be
    F(z)=\frac{A}{e^{z}-1+\gamma}
    \label{fz2}
    \ee
Eq.(\ref{fz2})
states that two events of magnitude $M_i$ and $M_j$ are correlated
over a characteristic time $\tau_{ij}$ and become independent when
$t_i-t_j
>\tau_{ij}$. As a consequence, only a small fraction of previous events
can affect the probability of future earthquakes so that, after a
certain time, Earth crust loses memory of previous seismicity.
This aspect is perhaps more realistic with respect to the idea,
contained in a power law correlation, that  events are all
correlated with each other and also gives rise to important implications
for seismic forecasting. The construction of seismic catalogs,
indeed, dates back to about 50 years, and according to
Eq.(\ref{fz2}) one can have good estimates of seismic hazard
without considering previous seismicity. This is no longer true if
one assumes a power law time decorrelation.
We want also to point
out that a general state-rate formulation \cite{Die} gives rise to
correlations between earthquakes that decay exponentially in
time. We finally observe, that taking into account only a fraction
of previous events in the evaluation of conditional probabilities, the
numerical procedure considerably speeds up.  In the case of long range
temporal correlations, CPU time grows with the number of events as
$N_e^2$, whereas in the case of an exponential tail the growth is
linear in $N_e$. For this reason, assuming the functional form
(\ref{fz2}) one can simulate very large sequences of events. In
particular for a different choice of parameters, one can
construct synthetic catalogs containing the same number of events
($ N_e=245000$ with $M \ge 1.5$) of the experimental
California Catalog. In Fig. \ref{fig3} we compare numerical and
experimental distributions $D(\Delta t,M_L)$ for three different
values of $M_L$. For each value of $M_L$, the
numerical curve reproduces the experimental data and
fulfills Eq.(\ref{cor}) (inset (a) in Fig.(\ref{fig3})). We have also 
evaluated the behaviour of $C(n)$ using the numerical catalog 
finding qualitative agreement with
experimental results (Fig.1). Finally, the 
numerical  magnitude distribution $P(M)$ fits very well
the experimental one (inset (b) in Fig.(\ref{fig3})).

After fixing $k$, we express numerical time unit in seconds and
we observe that the numerical catalog corresponds to a period of about
$9.9\ 10^9 sec \simeq 30$ years. 
Our model is therefore able to construct a
synthetic catalog covering about $30$ years, containing about
the same number of events and displaying the same statistical
organization in  magnitude and time of occurrence as real
California Catalog. The high efficiency of the model in
reproducing past seismicity indicates that the model is a good
tool for earthquake forecasting. In fact, given a seismic history,
Eq.(\ref{VJ}) together with Eq.s(\ref{scaling}, \ref{fz2})
gives the rate of occurrence of magnitude $M$ earthquakes at
time $t$ inside a considered geographic region.
We want to point out that similar extended analysis has never been
performed in seismicity. It is, indeed, the first time that a stochastic
model is able to produce a synthetic catalog with
all the features of the real experimental
intertime and magnitude distributions. Recently, Saichev and Sornette
\cite{sorn2} have calculated the intertime distributions with an analytical
approach to the ETAS models. Under the assumption that an earthquake can
have at most one first generation aftershock they find for the intertime
distribution a behaviour different from Eq.(\ref{cor}).

We finally observe that also spatial organization of seismic
events reveals some kind of scale invariance \cite{spatial,pacz,god}. This
indicates that also spatial distribution originates from a critical
behaviour of the Earth crust suggesting that a dynamical scaling
hypothesis as in Eq.(\ref{scaling}) can also work if one
appropriately introduces spatial dependencies. In this way it would be possible
to construct seismic hazard maps.

{\small Acknowledgments. 
This research was supported by EU Network Number
MRTN-CT-2003-504712, MIUR-PRIN 2004, MIUR-FIRB 2001.}

\end{document}